# PMU Data Feature Considerations for Realistic, Synthetic Data Generation

Ikponmwosa Idehen, Wonhyeok Jang and Thomas Overbye
Electrical and Computer Engineering
Texas A&M University
College Station, TX, USA
{iidehen, wjang777, overbye}@tamu.edu

*Abstract*—It is critical that the qualities and features of synthetically-generated, PMU measurements used for grid analysis matches those of measurements obtained from field-based PMUs. This ensures that analysis results generated by researchers during grid studies replicate those outcomes typically expected by engineers in real-life situations. In this paper, essential features associated with industry PMU-derived data measurements are analyzed for input considerations in the generation of vast amounts of synthetic power system data. Inherent variabilities in PMU data as a result of the random dynamics in power system operations, oscillatory contents, and the prevalence of bad data are presented. Statistical results show that in the generation of large datasets of synthetic, grid measurements, an inclusion of different data anomalies, ambient oscillation contents, and random cases of missing data samples due to packet drops helps to improve the realism of experimental data used in power systems analysis.

## I. INTRODUCTION

IN the era of synchronized power grid phasor measurements (or synchrophasors) sourced from fast-reporting, phasor measurement units (PMUs), the prospects of high-resolution, grid monitoring are ever increasing [1, 2]. A classification of the electric power system as a critical national infrastructure mandates that sensitive grid information, with the ability of revealing the true state of the system, be protected [3, 4]. The implication is the challenge posed to several research studies whose accessibility to original power systems data is hindered by strict confidentiality rules and requirements of non-disclosure agreements (NDAs) often associated with the utilization of these data. As an alternative, researchers often resort to the use of experimental data obtained from power system simulation studies. However, [3] stipulates that for an accurate operational assessment or long-range planning of the modern grid, it is essential that data of the right type and high-fidelity be utilized for research purposes.

The unique feature sets of reported, industry PMU measurements, as observed by the consistent variations in high-resolution, time-series measurements, can be attributed to several factors. Complex operation of the grid, for example, due to constantly-changing loads, control devices, operator actions, and several disturbances occurring on the system [5, 6]; and the influence of malfunctioning PMU ancillary components, such as drifting internal clocks, poor global positioning system (GPS) signal reception, low-accuracy instrument transformers, improperly-connected wires, and other measurement errors lead to the familiar fluctuations observed in PMU data [7-13].

The creation of existing synthetic networks [14-18] aim to generate artificial data for the purpose of research activities. However these measurements are often devoid of actual PMU data attributes that are due to the already-mentioned phenomena. Mostly comprising of only the simulated system dynamics, these seemingly, error-free measurements do not bear true representations of PMU datasets, and could cause researchers to make inaccurate conclusions based on idealistic, experimental results. Synthetic data generation for research purposes have also been addressed in several fields related to software testing, machine learning, and social networks [19-22]. Most of these approaches utilize intelligent techniques, such as genetic algorithms, ensemble-based methods, R-programming, and rely on pre-defined models, patterns or random number generators to create artificial data. Due to several component and human operator interactions with the electric grid, power system measurements are unique as they embed underlying grid dynamics which reflect the state of the system. Hence, power system measurements are not random nor do they strictly follow any pre-defined pattern. In order to circumvent the reliance on power system component models, [22] proposed the use of an intelligent generative adversarial network (GAN) machine learning technique to synthesize a realistic PMU time-series measurement. A limitation with this method is its significant level of dependence on real data, and for which the confidentiality issues associated with accessing real data was the original motivation for synthetic data production. Furthermore, the ability to modify or make certain inclusions to features in the synthetic dataset, using the proposed method, may be limited since the artificial data is based on an original measurement. In situations where large-scale, multivariate datasets are required from multiple, geographically-dispersed sources, such that they embed all underlying system dynamics in addition to local behaviors, and simultaneously capture the intricate spatio-temporal relationships known to exist in electric grids [23], an inability of the current methods to train multiple real data while satisfying the above requirements would significantly restrict their implementation [24].

In this paper, an analysis of characteristic features observable in PMU measurements obtained from a publically-sourced, industry dataset is carried out for input considerations in the production of realistic, synthetic power systems PMU data. In the first step, random variations unique to PMU measurements are presented, after which a variability term is used to capture the different variation components inherent in real data. Furthermore, evaluations of real voltage magnitude, angle and frequency data measurements are performed to determine the extents of occurrence of several data anomalies and disturbances which contribute to PMU data variability.

## II. REAL AND SIMULATION PMU DATA

Fig. 1 (a) and (b) [13] respectively show 20 seconds duration of real (in red) and simulated (in blue) per unit voltage measurements obtained from PMUs operating with different nominal ratings during a generator outage. The real data is obtained from a publicly available dataset provided by [25] from 123 actual PMUs, while the simulation data is generated from a contingency analysis carried out on a synthetic Texas network comprising of 2,000 buses [17]. In both cases, grid operating frequency and PMU report rates are 60 Hz, and 30 samples per second respectively.

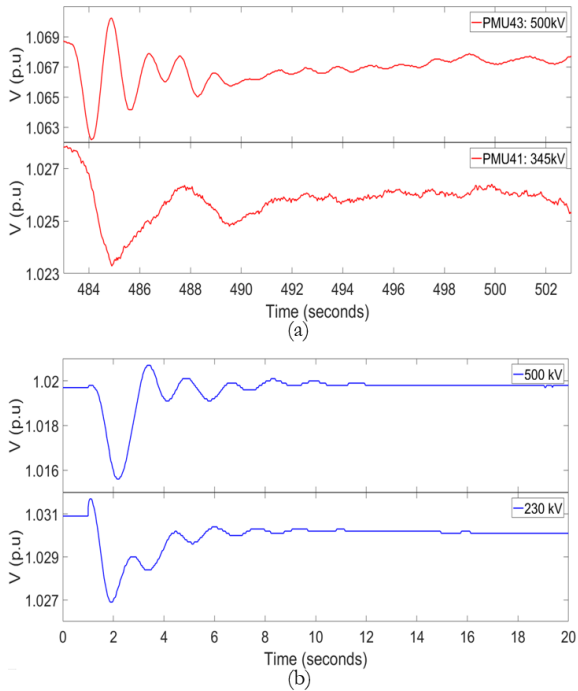

Fig. 1. 10-sec per unit voltages from real and simulation measurements

To ensure a fair comparison, the resolution on all voltage axis have been set to three decimal places. As observed, regardless of the nominal voltages of the nodes being monitored in the real system, trending voltage profiles are observed in both signals in Fig. 1 – indicative of the continuously changing state and dynamic interactions in a real grid. In contrast, Fig. 2 shows predominantly, flat voltage profiles observed in both simulation measurements, even after the occurrence of the generator event. When present, signal variations are few, less apparent and discontinuous, largely attributed to a preset, input time step used in the simulation during which grid states are evaluated. Depending on sensitivity levels of analytics where these measurements are used, these simulation data obtained from relatively inactive systems generate results which can exhibit significant difference from those which utilize real synchrophasor data obtained from steady, dynamics-driven, real power systems. This deviation of error-free, simulation data from actual synchrophasor measurements further becomes more apparent when, in addition to system dynamics, malfunctioning PMUs, limited network bandwidth and communication delays introduce data errors in real measurements increasing the inherent, signal variations.

## III. PMU MEASUREMENT VARIABILITY

According to [26], variation in grid voltage measurements can be represented as (1).

$$\sigma_{\Delta V_M}^2 = \sigma_{\Delta V}^2 + \sigma_\eta^2 \qquad (1)$$

$\sigma_{\Delta V}^2$ and $\sigma_{\Delta V_M}^2$ are the voltage signal variances before and after an introduced noise measurement variance, $\sigma_\eta^2$. Using any of the available techniques in the literature [8], $\sigma_\eta^2$ can be computed from the noise signal extracted from the original measurement, while $\sigma_{\Delta V}^2$ is evaluated as a by-product obtained from the filtered signal.

In order to factor in the existence of real data anomaly in actual PMU measurements (as will be discussed and observed in subsequent sections), (1) has been modified in this paper to incorporate an additional variance component, $\sigma_e^2$ as shown in (2).

$$\sigma_{\Delta V_M}^2 = \sigma_{\Delta V}^2 + \sigma_\eta^2 + \sigma_e^2 \qquad (2)$$

### A. Grid Dynamics Variability

The effect of pure grid dynamics is observed by performing an analysis of a segment of PMU measurement observed to possess predominantly noiseless and error-free data samples. Fig. 2 shows a 5-second segment of per unit voltage extracted from a 30-minute duration of measurements obtained from a PMU in the dataset [25].

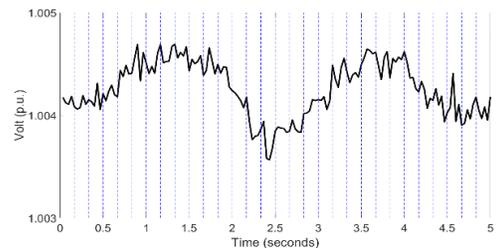

Fig. 2. 5-sec per unit voltage magnitude

The non-stationarity feature of the time-series data in Fig. 2 is observed by the continuous, seemingly-erratic state of the individual voltage samples. A high signal quality is captured by the three decimal place representation used in uncovering the inherent variation, and the high signal-noise ratio (SNR) value of 70 decibels computed for the first minute of measurement, which included the 5-second segment data. Research works, such as [8], show SNR of most power system measurements to be lower within a normal range of 41-47 decibels. Considering minimal data error, it can be concluded that the variation, $\sigma_{\Delta V_M}$ in Fig. 2 is approximately attributed to the sole contribution of

$\sigma_{\Delta V}$ as observed in (2). An analysis of the voltage profile in Fig. 2 reveals other hidden trends by observing its down-sampled signals. Fig. 3 shows average voltage (blue, dashed line) and corresponding variance (red, solid line) for every non-overlapping window comprising of every five data samples.

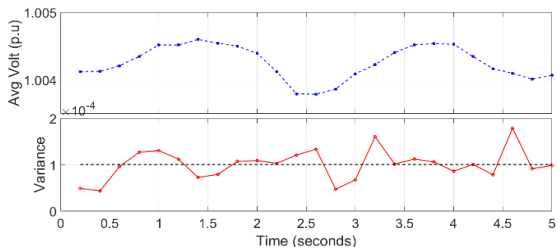

Fig. 3. Down-sampled: 5-sample window mean and variance

As observed, a moving-window average smoothens the original profile and exposes the underlying trend, while the variance reveals the extent of fluctuation among samples. An average variance of $10^{-4}$, hereafter referred to as an average variability in this paper, is also detected when a steady change in average voltage values is observed across all non-overlapping windows. Given that $\sigma_\eta$ and $\sigma_e$ are negligible, it can be concluded that this value of average variability provides a measure for the true fluctuation due to actual system dynamics interaction. However, it is important to note that in data segments where more random variations are associated with lower computed SNR values, this assumption is rendered invalid.

A further analysis of a real grid can be performed to reveal the underlying extent of system dynamics [13]. Fig. 4(a) and (b) show average variabilities and SNRs of the corresponding 5-second per unit voltage data segments for all 123 PMUs in the dataset.

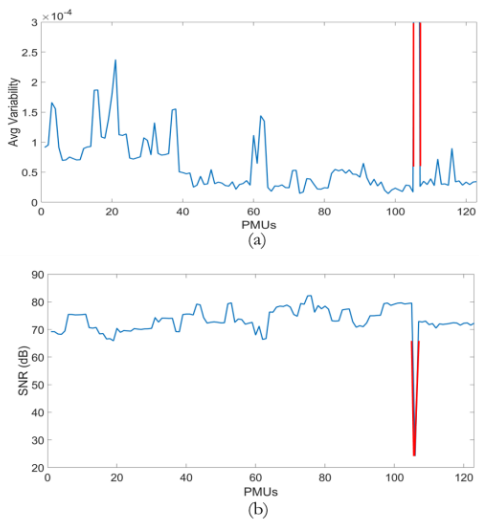

Fig. 4. (a) Average variability and; (b) SNR of all 123 real voltages

### B. Measurement Noise Variability

An unwanted disturbance, noise signals in power systems introduce additional variability to PMU data measurements. Generally not assumed to belong to a normal distribution [27], however power system noise are still oftentimes modeled as a normal distribution possessing a zero-mean and a standard deviation. Fig. 5 shows a one-minute duration of noise signal extracted from one of the PMU-reported frequency measurements in [25].

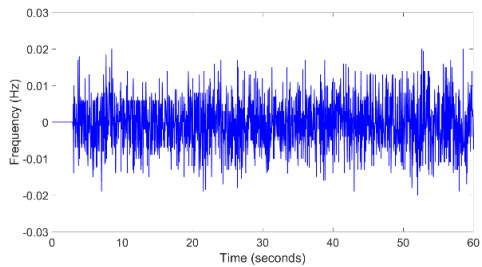

Fig. 5. Noise in 1-min frequency measurement

Zero frequency values observed in the first 3-seconds are attributed to the time interval during which no data filtering occurs when a moving-window, median filter [28] of order-90 was used to remove deviant data samples in the original measurement. The computed statistical properties of this noise signal is observed to be approximately zero mean and an SNR of 43.1 decibels typical of most power system measurements.

A further assessment of this signal to confirm the 'independent and identically distributed ($i.i.d$)' property of the noise signal samples can be carried out by computing an autocorrelation coefficient, $\rho$ as a function of different values of lag $k$ [29, 30].

$$\rho_k = \frac{E[(z_t - \mu)(z_{t+k} - \mu)]}{\sqrt{(E[(z_t - \mu)^2]E[(z_{t+k} - \mu)^2])}} \quad (3)$$

$z$ is a measurement sample; $\mu$ is a constant mean over the entire range of measurements; $E[(z_t - \mu)(z_{t+k} - \mu)]$ is an auto-covariance function which measures the co-variance between any sample that are $k$ distance apart; and $E[(z_t - \mu)^2]$ is a self-correlation of a sample (i.e., correlation with respect to itself). Fig. 6 is the generated autocorrelation function (ACF) for $k$ values up to 20.

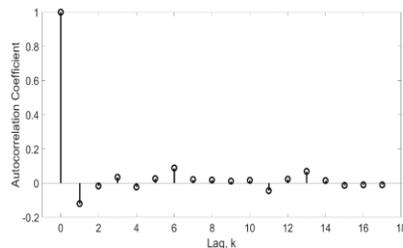

Fig. 6. Autocorrelation function for power systems noise signal

In an ideal scenario, noise samples are independent of each other, such that at any lag, $k \neq 0$, the ACF should equal zero, and one if otherwise. Fig. 6 approximates this behavior with a small ACF value of less than 0.2 at $k = 1$ before rolling off to zero at the next lag value. Neglecting its self-correlation (i.e., $k = 0$), an average ACF value of 0.06 over all the different lag values is indicative of the strong $i.i.d$ feature in this noise signal.

### C. Data Anomaly Variability

Instances of anomalous data samples are common features which contribute to PMU measurement variability. These data errors manifest in different forms ranging from errors due to malfunctioning sensor devices, time source errors, outlier samples due to transient grid events, and missing samples. Fig.7 shows 1-min, real voltage angle measurements obtained from five PMUs based on a reference angle for the entire system.

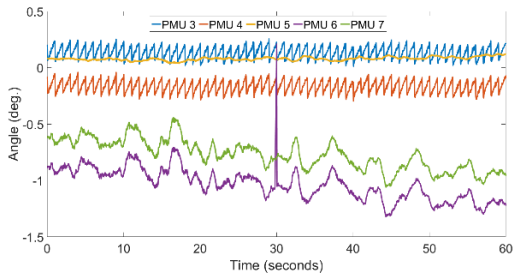

Fig. 7. 1-min voltage angle measurements

Neglecting the slow, and true trending angle measurement samples of PMU 5, the other PMUs are a reflection of some of the several, anomalous features which can occur in industry-obtained data, and must be handled prior application usage. The observations in Fig. 7 are thus: PMUs 3 and 4 exhibit time-skew errors due to clock drift errors [9, 12]; low frequency oscillations are observed in PMUs 6 and 7; and an outlier data point in PMU 6. In simulated, error-free voltage angles, transitions between measurement samples are, if any, smooth, and lacking of any of the above attributes thus, causing them to differ from industry data. Additionally, the extent of measurement variability, due to data anomaly, is affected by the frequency of occurrence of local outlier measurements which are often caused by dynamics such as the transient nature of line or capacitor switching, noise, extra-terrestrial effects of sharp climate change (e.g. lightning) which introduce significant deviations in data samples.

Fig. 8(a) and (b) show percentage outliers obtained from the analysis of outlier samples in real voltage magnitude (V.M) and real voltage angle (V.A) measurements in all 123 PMUs over a 30-minute duration of data from [25]. That is, each PMU reports 54,000 data samples.

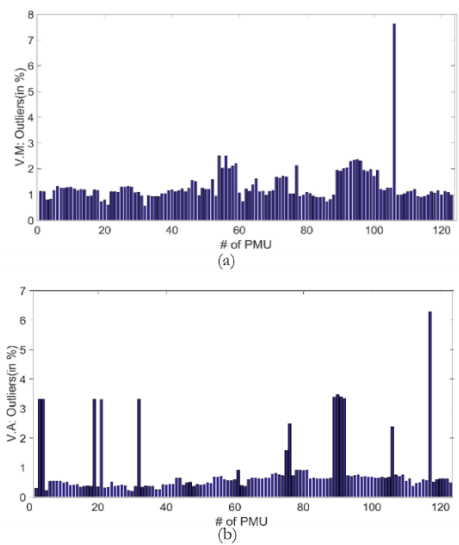

Fig. 8. Percentage outliers in (a) 30-mins V.M. (b) 30-mins V.A.

Both figures indicate consistent presence of outliers in power system measurements. Neglecting the PMU with over 7% V.M outliers, an average of 1%, corresponding to 540 samples, are noted to be outside their local vicinities. Significant outlier angle measurements in Fig. 8(b) is a unique feature of real voltage angle measurements where sporadic values of large deviations (as high as 100 deg.) need to be eliminated prior to data processing. While working directly with angle measurements can be challenging as a result of its significant non-stationarity, a first-order angle difference is used to obtain a stationary time-series for voltage angle outlier analysis. Here, the difference between consecutive voltage angles for any PMU is used to obtain a new time-series. That is,

$$VA_{d,i} = VA_{i+1} - VA_i; i = 1,2, \ldots 53999 \qquad (4)$$

$[VA_{d,i}]$ is the new, stationary, angle-difference time-series for a given PMU. Fig. 9 shows angle outliers in two selected PMUs with significant angle outliers [13]. For the entire 30-minute duration, a high amount of noisy signal is observed in PMU 106, while instances of large, random outliers are noted in PMU 76.

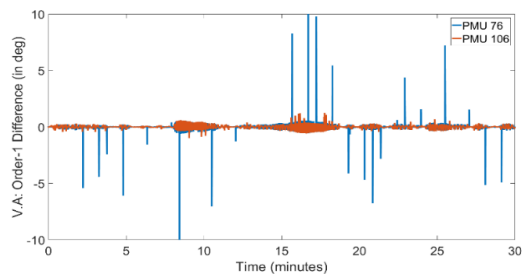

Fig. 9. First-order, stationary, 30-min voltage angles

## IV. PMU MISSING MEASUREMENTS

The underlying PMU network is prone to failures such as latency, time delay or actual communication link failures thus resulting in dropped data packets, and hence missing data samples. Data fillers, often identified in measurements as Not-a-Number (or NaN), or arbitrary values such as 9999 or -9999 are used to replace these missing sample points to ensure continuous operation of the PMU data reporting infrastructure.

A data completeness problem, missing PMU data samples can be quantified in terms of a drop-out rate ($\rho$) or a maximum drop-out (or gap) size ($\chi$) [7]. $\rho$, and shown in (5), quantifies the number of packets dropped in a time period, and $\chi$ is the largest contiguous set of time points during which no actual data is available.

$$\rho = \frac{\text{Number of dropped samples}}{\text{Total number of expected samples within time period}} \qquad (5)$$

An actual measure of frequency of occurrence of missing PMU data samples in a real scenario can be determined by assessing the 30-minute duration of data from [25]. Fig. 10(a) and (b) show the computed drop-out rates and gap sizes for all 123 PMUs respectively.

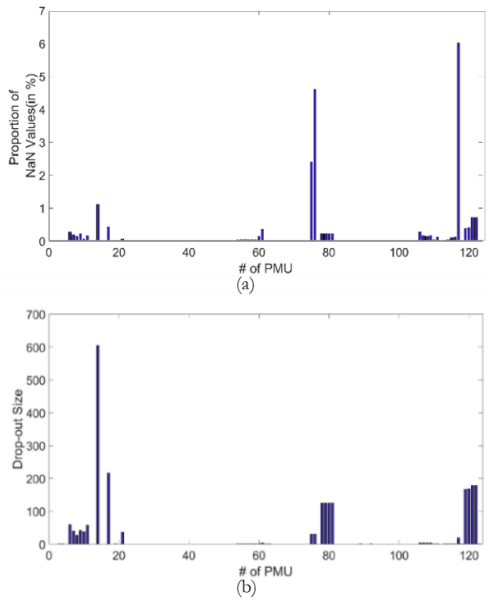

Fig. 10. (a) Drop-out rates and; (b) Maximum gap size in 123 PMUs

Three PMUs (with IDs 75, 76 and 117) in Fig. 10(a) are observed to show significant levels of missing data (i.e., beyond 2%), while Fig. 10(b) reveals 47 PMUs, accounting for 38% of all PMUs, had at least one instance of missing value. Fig. 10 (b) also indicates the absence of actual measurements for a significant time duration (i.e., > 3sec) in ten PMUs.

These statistics demonstrate the prevalent case of missing data samples, which define some of the unique features observed in actual PMU measurements. In real-life scenarios, engineers will often be faced with the problem of resolving missing samples prior to grid analysis.

## V. AMBIENT OSCILLATIONS

A consequence of multiple grid activities arising from automatic generator controls, switching actions of system components and constantly-changing system loads is the presence of hidden, low-frequency (LF) disturbance signal modes when small-signal stability procedures are carried out [31-33][30-32][29-31]. Real power grids are characterized by these ambient oscillations which embed the fundamental operating dynamics of their systems. Fig. 11 shows a one minute interval of real frequency measurements obtained from all 123 PMUs during which a disturbance was observed in the public dataset [25].

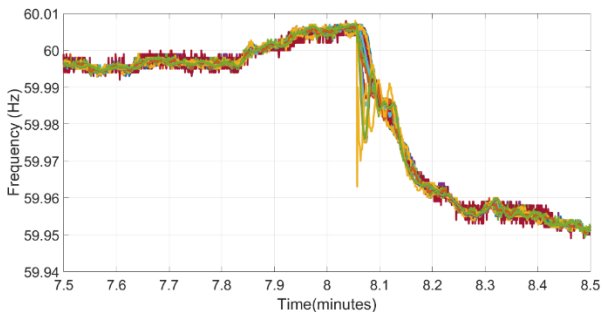

Fig. 11. First-order, stationary, 30-min voltage angles

Besides the significant disturbance of 8.05-mins during which a generator trip event occurs, other relatively lower disturbances are observed in the system. A choice of any of the frequency analysis techniques in the literature applied in multiple, overlapping time windows can help to reveal the inherent low-frequency oscillatory modes in the system. Fig. 12 shows frequencies, magnitudes and damping factors of only significant, oscillatory signal modes detected when a matrix pencil (MP) modal analysis was performed on 1-min frequency dataset using a moving-window width of 10-sec, and a step size of 5-sec.

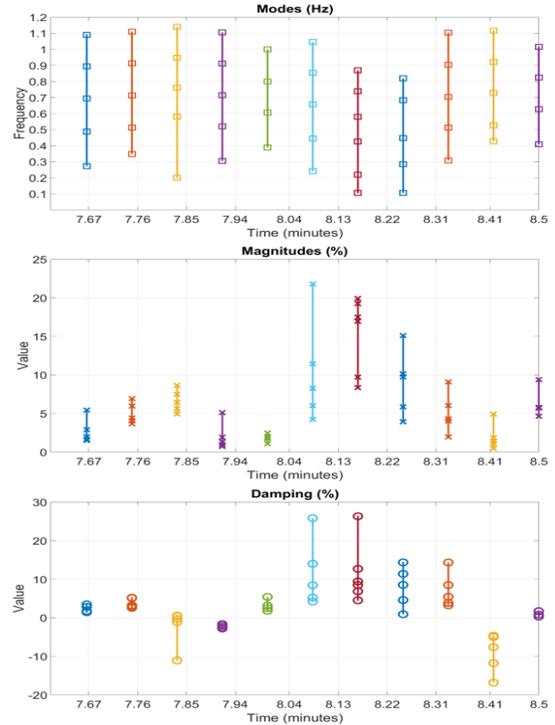

Fig. 12. First-order, stationary, 30-min voltage angles

A generator trip event is flagged by the high concentration of a large mix of significantly low frequency modes dataset observed between 8.04 and 8.31 min. This disturbance is characterized by dominant modes of 0.11 Hz and 0.22-0.28 Hz, which are then observed to diminish in strength in subsequent time windows apparently due to system damping.

With exception to the disturbance windows, frequency modes, greater than 0.9 Hz are observed in other windows. These are oftentimes attributed to the effects of local generator control oscillations, excitation and DC controls, electromechanical inter-area or local oscillations (between 0.15 – 1.0 Hz) are usually of more interest since they are mostly associated with system disturbances. However, the most significant observation is the presence of two system modes of approximately 0.3 Hz and 0.5 Hz which appear in all data windows. While very few time windows may indicate instances of under-damping of these modes, their relative high magnitudes in all windows indicate their dominant presence in the system. As part of an interconnected system, we conjecture that these signals constitute signature or ambient, system electromechanical modes [31] from which the PMU dataset has been obtained. These unique system modes will usually be attributed to the random input variations (e.g. constantly-changing load) and other generator dynamics specific to that system.

VI. CONCLUSION

Current methods generate experimental data from several power grid simulations, and utilize them directly in grid assessments. However, they are often devoid of those features, such as data variability similar to real dynamics, data error inclusions and missing samples, which define the unique qualities of PMU-derived measurements. In this paper, significant features of actual PMU data have been discussed for future consideration in the creation of large datasets of realistic, synthetic measurements. The prevalence of outlier measurements and missing data samples, through observed statistics from a publically, sourced industry dataset has been discussed. In addition, the presence of inherent, low-frequency oscillations for inclusion in synthetic measurements in order to define an underlying dynamics similar to an actual system has also been discussed.

It is critical that these observed features be incorporated in the production of synthetic data to improve their realism. This will ensure that researchers utilize realistic, synthetic data with comparable attributes and qualities of industry data in order to arrive at conclusions similar to those expected in real scenarios.

Future works are currently aimed at producing realistic PMU datasets through the application of industry-type, generation and load variation patterns on large-scale, synthetic networks, and thereafter infused with PMU data features.

ACKNOWLEDGEMENTS

The authors would like to thank the Power System Engineering Research Center (PSERC) under project T-57 for supporting this work.